\documentclass[aps,prc,twocolumn,10pt,superscriptaddress,preprintnumbers,showpacs,nofootinbib,a4paper,]{revtex4-1}

\usepackage{verbatim}
\usepackage[latin1]{inputenc}
\usepackage{graphicx,amssymb} 

\usepackage{dcolumn}

\usepackage{color}%

\graphicspath{{.}{../Figures-EPS/}}

\newcommand{\hzdr}{\affiliation{Helmholtz-Zentrum Dresden-Rossendorf (HZDR), Bautzner Landstr. 400, 01328 Dresden, Germany}}
\newcommand{\tu}{\affiliation{Technische Universit\"at Dresden, Dresden, Germany}}
\newcommand{\gsi}{\affiliation{GSI Helmholtzzentrum f\"ur Schwerionenforschung, 64291 Darmstadt, Germany}}
\newcommand{\pd}{\affiliation{Istituto Nazionale di Fisica Nucleare (INFN), Sezione di Padova, I-35131 Padova, Italy}}
\newcommand{\atomki}{\affiliation{MTA ATOMKI, H-4028 Debrecen, Hungary}}
\newcommand{\vkta}{\affiliation{Verein f\"ur Kernverfahrenstechnik und Analytik Rossendorf (VKTA), 01328 Dresden, Germany}}
\newcommand{\villigen}{\affiliation{Paul-Scherrer-Institut, CH-5232 Villigen, Switzerland}}

\begin{document}

\title{The resonance triplet at $E_\alpha = 4.5$\,MeV in the $^{40}$Ca($\alpha$,$\gamma$)$^{44}$Ti reaction} 

\thispagestyle{empty}

\author{Konrad Schmidt}\hzdr\tu 
\author{Shavkat Akhmadaliev}\hzdr 
\author{Michael Anders}\hzdr\tu 
\author{Daniel Bemmerer}\email{d.bemmerer@hzdr.de}\hzdr
\author{Konstanze Boretzky}\gsi 
\author{Antonio Caciolli}\pd 
\author{Detlev Degering}\vkta 
\author{Mirco Dietz}\hzdr\tu
\author{Rugard Dressler}\villigen
\author{Zolt\'an Elekes}\hzdr
\author{Zsolt F\"ul\"op}\atomki 
\author{Gy\"orgy Gy\"urky}\atomki 
\author{Roland Hannaske}\hzdr\tu 
\author{Arnd R. Junghans}\hzdr 
\author{Michele Marta}\hzdr\gsi 
\author{Marie-Luise Menzel}\hzdr\tu 
\author{Frans Munnik}\hzdr
\author{Dorothea Schumann}\villigen 
\author{Ronald Schwengner}\hzdr 
\author{Tam\'as Sz\"ucs}\atomki 
\author{Andreas Wagner}\hzdr 
\author{Dmitry Yakorev}\hzdr\tu
\author{Kai Zuber}\tu

\begin{abstract}
The $^{40}$Ca($\alpha$,$\gamma$)$^{44}$Ti reaction is believed to be the main production channel for the radioactive nuclide $^{44}$Ti in core-collapse supernovae. Radiation from decaying $^{44}$Ti has been observed so far for two supernova remnants, and a precise knowledge of the $^{44}$Ti production rate may help improve supernova models. The $^{40}$Ca($\alpha$,$\gamma$)$^{44}$Ti astrophysical reaction rate is determined by a number of narrow resonances. Here, the resonance triplet at $E_\alpha = 4497$, 4510, and 4523\,keV is studied both by activation, using an underground laboratory for the $\gamma$~counting, and by in-beam $\gamma$~spectrometry. The target properties are determined by elastic recoil detection analysis and by nuclear reactions. The strengths of the three resonances are determined to $\omega\gamma = (0.92\pm0.20)$, $(6.2\pm0.5)$, and $(1.32\pm0.24)$\,eV, respectively, a factor of two more precise than before. The strengths of this resonance triplet may be used in future works as a point of reference. In addition, the present new data directly affect the astrophysical reaction rate at relatively high temperatures, above 3.5\,GK. 
\end{abstract}

\maketitle

\section{Introduction}
\label{sec:Introduction}

\subsection{Astrophysics of $^{44}$Ti}

Core-collapse supernovae and their precursors, massive stars, are believed to be the sites of the weak astrophysical s-process and probably also the astrophysical r-process, significantly contributing to the production of the chemical elements heavier than iron \cite{Diehl11-Book}. 
Computer simulations have made progress in recent years in the description of these cataclysmic events \cite{Janka12-ARNPS}. However, much remains to be done, both from the observational and from the modeling point of view \cite{Burrows13-RMP}. 
While modern observatories have enabled the detection of far-away supernovae, for detailed studies one has to rely on objects that are located in the Milky Way. 

Radioactive isotopes may offer particularly rich insight~\cite{Diehl11-Book}. One such case is the radionuclide $^{44}$Ti, which decays by electron capture to $^{44}$Sc. It has a half-life of just $58.9\pm0.3$\,years~\cite{Ahmad06-PRC}, which may be somewhat increased in the ionized supernova environment \cite{Mochizuki99-AA}. Even still, $^{44}$Ti is emitted by young supernova remnants and thus offers a rare glimpse of nucleosynthesis in action. 

Using satellite-based $\gamma$-ray observatories, $^{44}$Ti has been detected in the Cassiopeia A supernova remnant: First by COMPTEL through the 1157\,keV $\gamma$ ray of its daughter $^{44}$Sc~\cite{Iyudin94-AA}, then also through the 68 and 78\,keV $\gamma$ rays of $^{44}$Ti itself, by the BeppoSAX~\cite{Vink01-ApJ} and INTEGRAL~\cite{Renaud06-ApJ} satellites. Combining these three independent measurements, the flux from $^{44}$Ti emissions in this line of sight has been determined with 12\% uncertainty~\cite{Renaud06-ApJ}. 

Very recently, the indirect conclusion that the decay of $^{44}$Ti powers the late light curve of the remnant of supernova~1987A~\cite{Jerkstrand11-AA} and illuminates its ejecta~\cite{Larsson11-Nature} has been directly confirmed by the observation of the 68 and 78\,keV $^{44}$Ti $\gamma$ rays, again with INTEGRAL~\cite{Grebenev12-Nature}. 
Surprisingly, the two supernova remnants Cas~A and SN1987A are the only cases with a confirmed detection of $^{44}$Ti, and in both cases the inferred amount of ejected $^{44}$Ti somewhat exceeds expectations from core-collapse supernova models. This leads to the question of whether $^{44}$Ti ejection is actually the exception rather than the rule for core-collapse supernovae~\cite{The06-AA}. However, before such far-reaching conclusions can be drawn, it is necessary to precisely understand the nuclear physics of $^{44}$Ti production. Nuclear reaction network calculations~\cite{The98-ApJ,Magkotsios10-ApJSS} have shown that $^{44}$Ti production is dominated by the $^{40}$Ca($\alpha$,$\gamma$)$^{44}$Ti reaction.

\subsection{Review of previous experiments}
\label{subsec:Review}

The 4.5\,MeV resonance triplet (Fig.~\ref{fig:Ti-44_levels}) in the $^{40}$Ca($\alpha$,$\gamma$)$^{44}$Ti reaction ($Q$ value 5.127\,MeV) was first studied by in-beam $\gamma$~spectroscopy using NaI(Tl) and Ge(Li) detectors \cite{Simpson71-PRC}. In a later work from the same group the individual resonance strengths were measured with 20\% uncertainty by Dixon {\it et al.}~\cite{Dixon80-CJP}. These initial experiments were mainly motivated by nuclear structure questions.

When $^{44}$Ti production in core-collapse supernovae moved to the foreground of considerations, a new series of experiments with a variety of techniques were reported. A pioneering new study was performed by Nassar {\it et al.} in inverse kinematics with a $^{40}$Ca beam incident on a helium gas cell~\cite{Nassar06-PRL}. The $^{44}$Ti produced was collected in a catcher and subsequently counted via accelerator mass spectrometry (AMS). Both a thin target, to determine the resonance strength of the 4.5\,MeV resonance triplet alone, and a thick target, to determine an averaged yield over a wide energy range, have been used. The thin target yield, dominated by the 4.5\,MeV resonance triplet studied here, was consistent with previous work (Table~\ref{table:LiteratureReview}). However, the thick target yield indicated a factor of three to five higher strength than the sum of resonance strengths found in the literature~\cite{Nassar06-PRL}.

The $^{40}$Ca($\alpha$,$\gamma$)$^{44}$Ti excitation function was then studied over a wide energy range in inverse kinematics by Vockenhuber {\it et al.}, using the DRAGON recoil mass spectrometer~\cite{Vockenhuber07-PRC}. The derived resonance strength for the present triplet was lower than Nassar {\it et al.}, but still consistent within the error bars (Table~\ref{table:LiteratureReview}). However, the derived reaction rate from the Vockenhuber {\it et al.} data was overall lower than Nassar {\it et al.} The discrepancy reaches a factor of two at high temperatures. 

In a subsequent experiment by Hoffman {\it et al.}, the $^{40}$Ca($\alpha$,$\gamma$)$^{44}$Ti yield was measured in direct kinematics, using infinitely thick CaO solid targets~\cite{Hoffman10-ApJ}. The integral yield was determined over a wide energy range, both by in-beam $\gamma$~spectroscopy and by activation, and was found to be significantly below theoretical expectations. As a consequence, the overall astrophysical reaction rate derived by Hoffman {\it et al.} was even lower than that of Vockenhuber {\it et al.}. Due to the large target thickness, no individual resonance strengths could be resolved.

\begin{table}
  \caption{Summed resonance strength of the triplet at $E_\alpha = 4.5$\,MeV, from this work and from the literature.}
  \begin{tabular}{D{.}{.}{1}@{$\,\pm\,$}D{.}{.}{1}ll}
    \hline
    \hline
    \multicolumn{2}{c}{$\omega\gamma$ [eV]}	& \multicolumn{1}{l}{Reference}	& \multicolumn{1}{l}{Technique} \\
    \hline
    8.3	& 1.7	& Dixon {\it et al.}~\cite{Dixon80-CJP}			& in-beam $\gamma$ spectroscopy \\
    8.8	& 3.0	& Nassar {\it et al.}~\cite{Nassar06-PRL}		& AMS \\
    7.6	& 1.1	& Vockenhuber {\it et al.}~\cite{Vockenhuber07-PRC}	& recoil detection \\
    9.0	& 1.2	& Robertson {\it et al.}~\cite{Robertson12-PRC}		& in-beam $\gamma$ spectroscopy \\
    8.4	& 0.6	& present work						& activation and \\
    \multicolumn{3}{c}{\,}						& in-beam $\gamma$ spectroscopy \\
    \hline
    \hline
  \end{tabular}
  \label{table:LiteratureReview}
\end{table}

Very recently, Robertson {\it et al.} studied the excitation function again over a wide energy range, using solid CaO targets and in-beam $\gamma$~spectroscopy with a 4$\pi$ NaI(Tl) summing crystal. Individual resonance strengths were derived from the whole excitation function using a fitting routine with the resonance energies and strengths as free parameters. The result for the 4.5\,MeV resonance triplet was somewhat higher, but consistent within error bars with the other previous works.  

\subsection{Aim of the present work}

Despite the wealth of available data (Sec.~\ref{subsec:Review}) and also some theoretical efforts~\cite{Rauscher00-NPA}, no consistent picture has emerged on this crucial nuclear reaction. The aim of the present work is to clarify some of these discrepancies by providing a precise normalization value that is accessible with relative ease to the experimentalist, and at the same time addresses the higher part of the astrophysically relevant energy range. 

\begin{figure}
 \includegraphics[width=1.0\columnwidth]{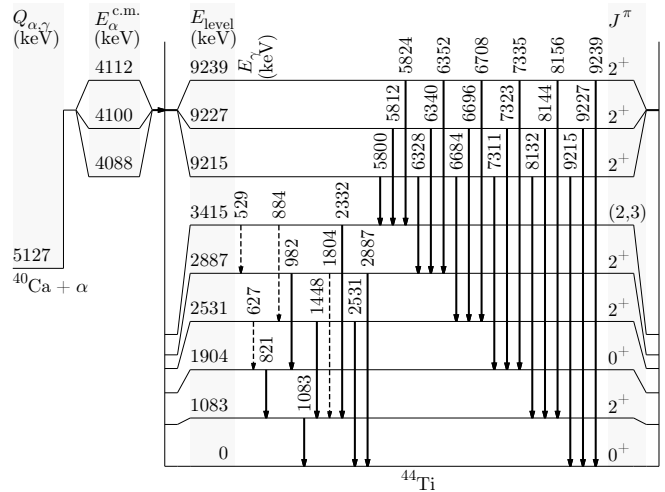}
 \caption{Reduced level scheme of $^{44}$Ti, showing only those levels and transitions that are known from the literature to be populated in the $^{40}$Ca($\alpha$,$\gamma$)$^{44}$Ti resonance triplet at $E_\alpha = 4.5$\,MeV, corresponding to center-of-mass energies of 4088, 4100, and 4112\,keV. Transitions that are observed (not observed) in the present work are shown with full (dashed) arrows. Level energies are given in horizontal orientation, $\gamma$-ray transition energies in vertical orientation. All energies in keV.}
 \label{fig:Ti-44_levels}
\end{figure}

To this end, the resonance triplet at $E_\alpha = 4.5$\,MeV in the $^{40}$Ca($\alpha$,$\gamma$)$^{44}$Ti nuclear reaction (Fig.~\ref{fig:Ti-44_levels}) is studied in a precision experiment, both by activation and by in-beam $\gamma$-ray spectroscopy. It is aimed to extend this study to  several lower-energy resonances in the future.

\section{Experiment}\label{sec:Experimental}

The irradiations have been carried out at the 3\,MV~Tandetron of Helmholtz-Zentrum Dresden-Rossendorf~\cite{Friedrich96-NIMA}. The accelerator provided a beam of $E_\alpha = 4.5$\,MeV $^4$He$^{++}$ ions with an intensity of 2.0\,--\,2.6\,$\mu$A (1.0\,--\,1.3 particle-$\mu$A) on target.

\subsection{Setup used for the irradiations}
\label{subsec:Setup}

\begin{figure}
 \includegraphics[width=1.0\columnwidth]{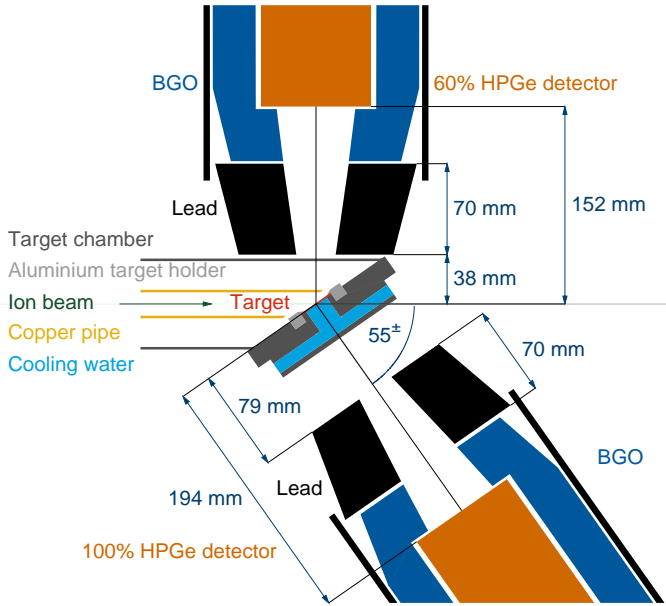}
 \caption{(Color online) Schematic view of the experimental setup used for the irradiations.}
 \label{fig:setup_scheme}
\end{figure}

The beam from the accelerator is bent by 30$^\circ$, transported through a drift tube including electrostatic quadrupoles and horizontal and vertical deflector units and then enters the target chamber (Fig.~\ref{fig:setup_scheme}). 

At the entrance of the target chamber, the beam spot is limited by a water cooled collimator with a circular opening of 5\,mm diameter. The current on the collimator is 10\,--\,30\% of the target current. The beam then passes through a copper tube of 30\,mm diameter that extends up to 2\,mm from the target surface and is biased to -100\,V in order to force secondary electrons back to the target. The beam current was measured with a precision current integrator, and the charge pulses were recorded both in the list mode data acquisition system and by an analog scaler unit. An uncertainty of 1\% is estimated for the resulting beam intensity. 

The target consists of calcium hydroxide deposited on a 0.22\,mm thick tantalum backing (Sec.~\ref{subsec:TargetPreparation}). The tantalum backing was directly cooled by deionized water, ensuring a constant temperature during the irradiations. The beam line and the target chamber were kept under high vacuum during the irradiations. Near the target, a pressure of 2\,--\,$7\cdot10^{-7}$\,mbar was measured. 

\subsection{Detection of in-beam $\gamma$ rays}
\label{subsec:GammaDetection}

The prompt $\gamma$ rays emitted during the irradiations were detected by two escape-suppressed high-purity germanium (HPGe) detectors of 100\% and 60\% relative efficiency that were placed at 55$^\circ$ and 90$^\circ$ with respect to the beam axis, respectively. The escape suppression was achieved by veto detectors of 3\,cm thick bismuth germanate (BGO) crystals. The veto signals were timestamped and logged in the list mode data acquisition system, so that vetoed and unvetoed histograms could be extracted in the offline analysis. 

The signals from the two HPGe detectors were split and passed to two independent data acquisition chains. In the first chain, the preamplifier outputs were directly passed to the 100\,MHz, 14-bit CAEN N1728B analog to digital converter. The trigger was generated inside the unit, using a digital trigger algorithm. Triggers that fell within the integration window of the preceding event were counted but not used for event generation. Inside the unit, the pulse shapes were digitized, integrated, and timestamped using a moving window deconvolution algorithm. The difference between the total trigger count and the number of converted events was used to estimate the dead time of the system. The second data acquisition chain recorded histograms of the amplified and escape-suppressed HPGe systems in an Ortec 919E ADC unit with Gedcke-Hale deadtime correction. Dead time corrections of up to 7\% for the helium beam runs and up to 24\% for the hydrogen beam runs had to be considered.

\begin{figure}
 \includegraphics[width=1.0\columnwidth]{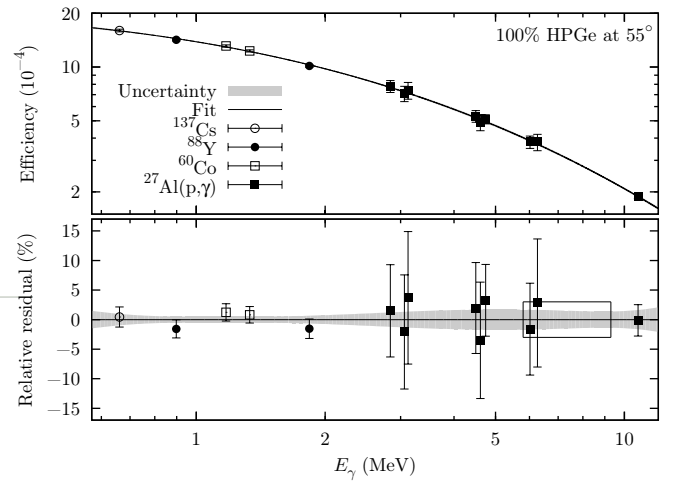}
 \caption{Top panel: $\gamma$-ray detection efficiency data and parameterization for the 55$^\circ$ detector. Bottom panel: Relative residual of the parameterization. The assumed 3\% uncertainty for the region of the primary $\gamma$ rays is shown as dashed rectangle.}
 \label{fig:efficiency}
\end{figure}

The absolute $\gamma$-ray detection efficiency was determined using $^{137}$Cs, $^{60}$Co, and $^{88}$Y activity standards provided by Physikalisch-Technische Bundesanstalt (PTB) with a 1$\sigma$ activity uncertainty of 0.5\%. The efficiency curve was then extended to high energy (Fig.~\ref{fig:efficiency}) based on a calibration run performed previously in the same setup, using the $E_{\rm p} = 992$\,keV resonance of $^{27}$Al(p,$\gamma$)$^{28}$Si. The ratio of emission rates of the 1779 and 10764\,keV $\gamma$ rays is well known~\cite{Zijderhand90-NIMA}, and the $^{27}$Al(p,$\gamma$)$^{28}$Si data were used in a relative manner. For the case of the 90$^\circ$ detector, also a correction for the angular distribution~\cite{Anttila77-NIM} had to be included. The previously determined branching ratios and angular distributions for several other  $^{27}$Al(p,$\gamma$)$^{28}$Si $\gamma$ rays~\cite{Anttila77-NIM} were also used. The resulting uncertainty for the efficiency is estimated to be 3\% for 5.8\,--\,9.3\,MeV (Fig.~\ref{fig:efficiency}), the energy region of the primary $\gamma$ rays from the $^{40}$Ca($\alpha$,$\gamma$)$^{44}$Ti reaction.

\subsection{Target preparation and handling}
\label{subsec:TargetPreparation}

The targets have been evaporated from highly pure\footnote{Merck 99.95 suprapur.} CaCO$_3$ on top of 0.22\,mm thick circular tantalum disks of 27\,mm diameter. The material was not isotopically enriched, in order to limit its fluorine contamination. When heated, CaCO$_3$ loses CO$_2$ and forms CaO, which may later turn into Ca(OH)$_2$ when reacting with ambient water vapour. A typical target thickness of 36\,$\mu$g/cm$^2$ was reached. After production, the targets were stored under inert gas in order to exclude any water vapour and transported to the experimental site. 

The targets were exposed to regular air for about 10\,min during the mounting and dismounting procedure. After mounting, the chamber was promptly evacuated to $10^{-7}$\,mbar. After dismounting, the targets were again stored in an airtight chamber filled with dry nitrogen. 

After the irradiation was concluded, the activated samples were inserted into a double holding ring structure that allowed safe handling without touching the surface. The activated targets were then brought to the Felsenkeller underground $\gamma$-counting facility, where their activity was determined (Sec.~\ref{subsec:ActivityDetermination}). 

\subsection{Stopping power and effective stopping power}
\label{subsec:TargetAnalysis}

In order to correctly determine the resonance strength for the $^{40}$Ca($\alpha$,$\gamma$)$^{44}$Ti reaction, the effective stopping power $\varepsilon_{\rm eff}^{\rm He}(E)$ must be known. This parameter dominates the systematic uncertainty and therefore warrants a detailed discussion. According to Bragg's rule, $\varepsilon_{\rm eff}^{\rm He}(E)$ is given by the stopping power of helium ions per $^{40}$Ca nucleus in the target:
\begin{equation}\label{eq:EffectiveStoppingPower_He}
\varepsilon_{\rm eff}^{\rm He}(E) = 
\frac{1}{\eta_{40}} \cdot \left( \varepsilon_{\rm Ca}^{\rm He}(E) + x \cdot \varepsilon_{\rm O}^{\rm He}(E) + y \cdot \varepsilon_{\rm H}^{\rm He}(E) \, \right)
\end{equation}
For the purpose of this relation, the target is assumed to be of stoichiometry CaO$_x$H$_y$. 
$\varepsilon_{\rm Ca}^{\rm He}(E)$ is the stopping power of helium ions in calcium,
$\varepsilon_{\rm O}^{\rm He}(E)$ in solid oxygen, and
$\varepsilon_{\rm H}^{\rm He}(E)$ in solid hydrogen.
The isotopic ratio of $^{40}$Ca in natural calcium is $\eta_{40} = (96.94\pm0.03)$\%. The bandwidth of abundances observed in natural materials~\cite{Coplen02-PAC} is used as a very conservative error bar for $\eta_{40}$.

The stopping powers $\varepsilon_{\rm O}^{\rm He}(E)$ and $\varepsilon_{\rm H}^{\rm He}(E)$ of helium ions in oxygen and hydrogen are experimentally well known and show a scatter of 2.0\% and 3.3\%, respectively, with respect to the general fit curve of SRIM~\cite{Ziegler10-NIMB}. There are no experimental data on helium ion stopping in calcium, so the general SRIM uncertainty of 3.5\% for helium ions~\cite{Ziegler10-NIMB} is adopted here as the uncertainty for $\varepsilon_{\rm Ca}^{\rm He}(E)$. For hydrogen ions, the stopping powers are known with 2.9\% (oxygen), 2.8\% (hydrogen), and 2.1\% (calcium) uncertainty, respectively~\cite{Ziegler10-NIMB}. An independent evaluation of several different stopping power codes found a mean deviation between SRIM~\cite{Ziegler10-NIMB} results and proton beam stopping data of $(-0.6\pm3.8)$\% for 17~solid elements where a complete comparison was possible~\cite{Paul05-NIMB}. For helium ions, the analogous comparison gave a $(-0.1\pm3.3)$\% deviation~\cite{Paul05-NIMB}. Henceforth, the widely accepted SRIM~\cite{Ziegler10-NIMB} data and error bars are used for the present purposes.

Some deviations from Bragg's rule have been observed for compounds containing oxygen and hydrogen, in particular near the Bragg peak. The present energies are higher, so these effects are mitigated here but still need to be included. Using the so-called Core and Bond approach~\cite{Ziegler88-NIMB} for stoichiometric Ca(OH)$_2$, due to the two O--H bonds a correction factor of 0.983 is found for $\varepsilon_{\rm eff}^{\rm H}$(1.8\,MeV) and a factor of 0.997 for $\varepsilon_{\rm eff}^{\rm He}$(4.5\,MeV). 

Two different methods have been used  in order to determine the stoichiometric ratio $x$ for three different targets produced during the same process: the elastic recoil detection (ERD) technique for target \#30 (Sec.~\ref{subsec:ERD}), and nuclear reactions for targets \#31 and \#32 (Sec.~\ref{subsec:NuclearReactions}). 

\begin{figure}
  \includegraphics[width=1.0\linewidth]{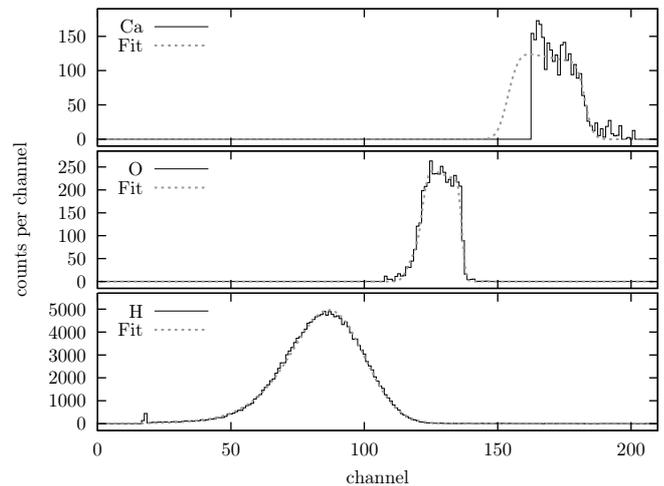}
  \caption{ERD results for target~\#30. Top and middle panel: Regions of carbon and oxygen recoils, respectively, projected on the energy axis. The lower-energy part of the calcium spectrum, corresponding to deeper target layers, cannot be analyzed due to overlap with backscattered copper ions and is therefore cut off at channel~162. Bottom panel: Spectrum from the PIPS detector for hydrogen ions.}
 \label{fig:ERDA}
\end{figure}

\subsection{Elastic recoil detection analysis}
\label{subsec:ERD}

For one target~(\#30) that was produced during the same process as the targets used for the irradiations, a complete ERD analysis  was performed (Fig.~\ref{fig:ERDA}). This sample was bombarded with copper ions of 50\,MeV kinetic energy from the 6\,MV Tandetron accelerator of HZDR at 15$^\circ$ grazing angle. The resulting recoils were detected in a Bragg-type ionization chamber (for boron and heavier ions) or a passivated implanted planar silicon (PIPS) detector preceded by a 18\,$\mu$m aluminum foil (for hydrogen ions), in a setup described previously~\cite{Kreissig98-NIMB}. The copper beam energy was sufficient to separate most of the calcium counts from the kinematic locus of the elastically backscattered copper ions. After gating for calcium or oxygen, the data were then projected on the energy axis and analyzed using the NDF software version 9.3g~\cite{Barradas97-APL}. No gating was necessary for the PIPS spectrum which was dominated by hydrogen ions, because the scattered copper ions and most recoils are stopped by the foil.

As a result of the ERD analysis, three layers have been obtained. The first one of 120\,nm thickness (210\,nm seen by a proton beam that is incident at 55$^\circ$ with respect to the target normal) had an elemental ratio of ${\rm Ca:O:H}=1:(1.79\pm0.20):(1.97\pm0.22)$, consistent with calcium hydroxide, Ca(OH)$_2$. The error bar for the O/Ca and H/Ca ratios is dominated by the 10\% uncertainty on the number of detected calcium recoils. The remaining two layers of this target description are first an interlayer containing tantalum and oxygen, and then the 0.22\,mm thick backing of pure tantalum. 

\begin{figure}
  \includegraphics[width=1.0\linewidth]{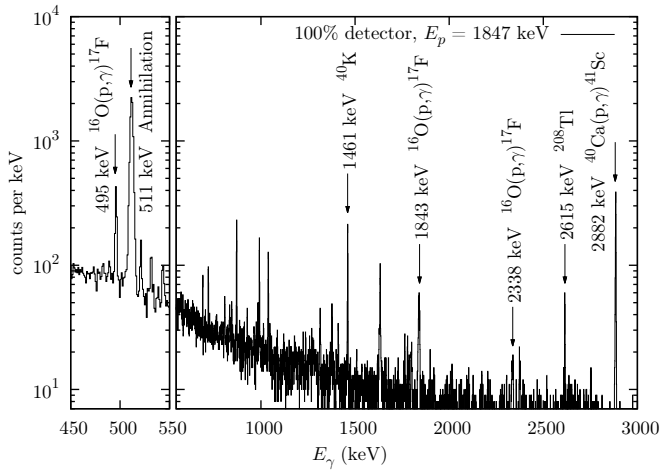}
  \caption{In-beam $\gamma$-ray spectrum on top of the plateau of the $E_{\rm p} = 1.842$\,MeV $^{40}$Ca(p,$\gamma$)$^{41}$Sc resonance. The $\gamma$ lines used for the target analysis and the main background lines have been labeled.}
 \label{fig:ProtonScanSpectrum}
\end{figure}

Various other configurations were tested in the fit software, however a reasonable fit was only reached for a target description consistent with the above numbers for the layer containing calcium. The depth resolution of the hydrogen profile is poor due to energy straggling in the foil before the PIPS detector. Even still, target descriptions where the hydrogen was contained only in the tantalum backing could be ruled out by the energetic position of the hydrogen peak. The ratio O/Ca was found to be consistent with 2 at all depths where calcium was observed. 

Finally, from the average of $x$ and $y$ a stoichiometric ratio of $1:(1.88\pm0.21)$ for ${\rm Ca:(OH)}$ is adopted here, resulting in a total areal density of $(7.8\pm0.4)\cdot10^{17}$\,at/cm$^2$ oxygen atoms (at 55$^\circ$ orientation) from the ERD data. 

\subsection{Target analysis by nuclear reactions}
\label{subsec:NuclearReactions}

The stoichiometric ratio was determined by nuclear reactions for targets \#31 and \#32. The amount of oxygen in the target was determined from the yield of the DC$\rightarrow$495 direct capture $\gamma$ rays from the $^{16}$O(p,$\gamma$)$^{17}$F nuclear reaction, observed  at $E_{\rm p} = 1.85$\,MeV (near $E_\gamma$ = 1843\,keV, spectrum in Fig.~\ref{fig:ProtonScanSpectrum}). The cross section of this process is known to 6\% precision in the relevant energy range~\cite{Mohr12-NIMA}, and the angular distribution for this transition has been measured previously in two independent experiments with consistent results~\cite{Rolfs73-NPA_SpecFac,Chow75-CJP}. 

Two other $\gamma$ rays are emitted from the $^{16}$O(p,$\gamma$)$^{17}$F reaction, as well, and also observed here. However, the DC$\rightarrow$GS line (near $E_\gamma$ = 2338\,keV) is weak, and the angular distribution for the 495$\rightarrow$GS $\gamma$-ray transition is not known from experiment. The isotopic enrichment of $^{16}$O in natural oxygen (99.76\%) is stable in all natural compounds on the level of 0.02\%~\cite{Coplen02-PAC}, so $^{16}$O is a proper tracer for all stable isotopes of oxygen. The areal density of oxygen atoms, in units of 10$^{17}$\,at/cm$^2$, is found by this procedure to be $8.9\pm1.1$ for target \#31 and $8.8\pm1.1$ for target \#32, consistent with the ERD result for target \#30 (Sec.~\ref{subsec:ERD}). 

\begin{figure}
 \includegraphics[width=1.0\linewidth]{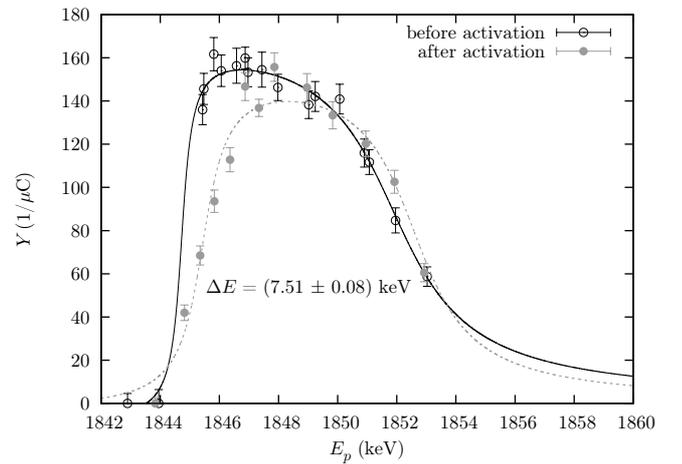}
 \caption{Proton beam scans of target \#32 before and after the activation, using the $E_{\rm p} = 1.842$\,MeV resonance in the $^{40}$Ca(p,$\gamma$)$^{41}$Sc reaction. The yield of the 2882\,keV $\gamma$ ray is plotted as a function of proton beam energy. The lines are plotted to guide the eye.}
 \label{fig:ProtonScanProfile}
\end{figure}

In order to determine the stoichiometric ratio $x$ without using the ERD result, not only the amount of oxygen atoms, but also the beam energy loss inside the target, i.e. the energetic target thickness, must be known. This quantity was determined in two different ways. 

First, based on the line shape of the two $^{16}$O(p,$\gamma$)$^{17}$F direct capture $\gamma$ rays, a proton beam energy loss of $7.62\pm0.19$ ($7.7\pm0.5$)\,keV was determined for target \#31 (\#32), respectively.

Second, the energetic thickness of targets \#31 and \#32 was determined from a scan of the $E_{\rm p} = 1.842$\,MeV resonance in the $^{40}$Ca(p,$\gamma$)$^{41}$Sc reaction. This narrow resonance emits $\gamma$ rays of 2.882\,MeV energy with 99.9\% branching ratio \cite{Zijderhand87-NPA}. The determined thickness was $7.35\pm0.14$ ($7.51\pm0.08$)\,keV for target \#31 (\#32), respectively, consistent with the result from the $^{16}$O(p,$\gamma$)$^{17}$F direct capture $\gamma$ lineshape.

Assuming the atomic concentrations of oxygen and hydrogen to be equal in targets \#31 and \#32, as shown in the ERD analysis for target \#30, one can then calculate the O/Ca ratio $x$ based on the given areal oxygen density and the energetic width of the target. The result is $x=1.9\pm0.6$ ($1.8\pm0.5$) for target \#31 (\#32), respectively (Fig.\,\ref{fig:ProtonScanProfile}), in good agreement with the ERD result for target \#30 and again consistent with calcium hydroxide.

Summarizing, the stoichiometry was determined by the ERD method for target \#30 and in a completely independent way based on proton-induced nuclear reactions for targets \#31 and \#32. The results for the three targets are mutually consistent. As they had been produced in the same process, the ERD result is adopted for the further analysis. 

It is not fully understood how the initial CaO transformed into Ca(OH)$_2$ during handling, irradiation, or storage. However, it is striking that the same stoichiometry is found for target \#30 (not exposed to helium or proton beam) and targets \#31 and \#32 (heavily irradiated with helium and proton beams). Therefore it seems reasonable to believe that the hydrogen and oxygen necessary to the CaO $\rightarrow$ Ca(OH)$_2$ transformation were provided by the tantalum backing.  

The proton resonance scan of target \#32 shows a shift of $\Delta$E$_{\rm p}$ = 1\,keV to higher energies. This value is consistent with the reproducibility of the accelerator energy calibration from day to day. However, such a shift may in principle also be caused by the buildup of an impurity layer on the target. Even if it exists, such a layer would change the stopping powers only negligibly.

\subsection{Monitoring of the irradiations}
\label{subsec:Monitoring}

The target can be expected to evolve under intense $^4$He$^{++}$ beam irradiation. Proton beam scans were performed before and after the $^4$He$^{++}$ activations, so the resulting target profile (Fig.~\ref{fig:ProtonScanProfile}) can be used to study the evolution of the target. The fresh target shows a sharply rising low-energy edge of the target and a flat plateau. The tapering off of the plateau towards high energies can be ascribed to energy straggling of the proton beam and is therefore not a sign of a non-rectangular target profile. After the irradiation, the plateau is shifted to higher energies, and the left edge is more gradual.

\begin{figure}
 \includegraphics[width=1.0\columnwidth]{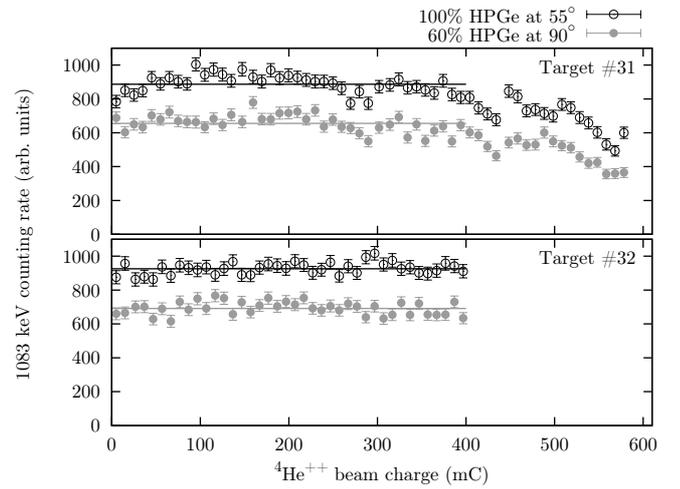}
 \caption{Rate of detected 1083\,keV $\gamma$ rays per He$^{++}$ dose, in the two HPGe detectors during the irradiation of targets \#31 (top panel) and \#32 (bottom panel). Open circles: detector at 90$^\circ$. Filled circles: detector at 55$^\circ$. The horizontal lines show the average counting rate up to a charge of 400\,mC.}
 \label{fig:CountingRate1083}
\end{figure}

During the He$^{++}$ irradiation, the stability of the target was monitored using the most intensive $\gamma$ ray from the $^{40}$Ca($\alpha$,$\gamma$)$^{44}$Ti reaction, the 1083\,keV $\gamma$ ray resulting from the decay of the first excited state of $^{44}$Ti (Fig.~\ref{fig:CountingRate1083}). 

During the irradiation of target \#31, after a period of constant counting rate up to a $^4$He$^{++}$ dose of 400\,mC, the rate began to degrade. The irradiation was stopped when the counting rate fell below 70\% of the original plateau value. Integrating under the yield curves, the  overall number of counts is reduced by 8.2\% due to this effect, so a correction of ($8.2\pm1.6$)\% has to be taken into account for this target, assuming a conservative value of 20\% relative uncertainty.

Target \#32 was irradiated for 400\,mC, and no decrease of the counting rate was observed at all (Fig.~\ref{fig:CountingRate1083}). 

\subsection{Activity determination in Felsenkeller}
\label{subsec:ActivityDetermination}

After irradiation, the samples were transported to the Felsenkeller underground $\gamma$-counting facility, which is shielded from cosmic-rays by a 47\,m thick rock overburden \cite{Szucs12-EPJA}. The samples were then placed on a specially designed sample holder in close geometry, at a distance of 1\,cm from the end cap of a p-type HPGe detector with 90\% relative efficiency \cite{Koehler09-Apradiso}. 

In order to determine the $\gamma$-ray counting efficiency in this geometry for the very weak $^{44}$Ti/$^{44}$Sc samples produced here, special calibration samples were used. Samples of an activity in the 100\,Bq range were produced in the ERAWAST project \cite{Dressler12-JPG} by evaporation of a $^{44}$Ti-containing aqueous solution on a tantalum backing of exactly the same size as those used for the targets (Sec.~\ref{subsec:TargetPreparation}), and letting the solvent evaporate. The activity of these $^{44}$Ti/$^{44}$Sc samples was then determined in far geometry at a distance of 20\,cm from a HPGe detector, based on $^{137}$Cs, $^{60}$Co, and $^{88}$Y activity standards. The actual $^{44}$Ti/$^{44}$Sc distribution on the calibration samples was measured with the image plate technique. As a result, the activity on these calibration samples was known with typically 1.5\% precision (see Ref.~\cite{Schmidt11-Diplom} for more details).  

\begin{figure}
 \includegraphics[width=1.0\linewidth]{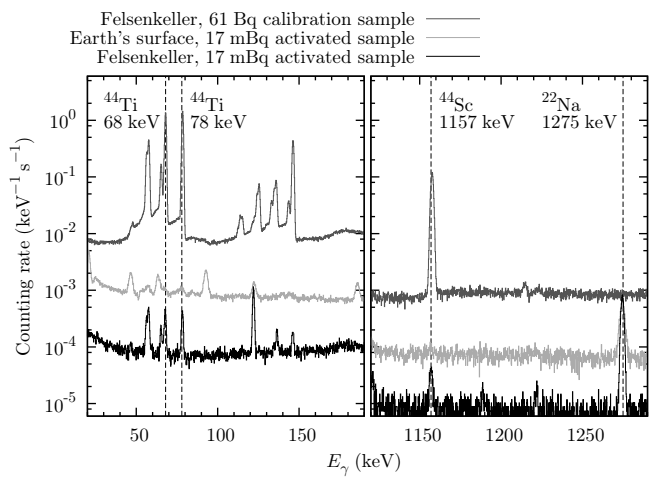}
 \caption{Decay spectra of $^{44}$Ti samples, measured in a low-background counting facility at the Earth's surface and in the ultra-low-background facility Felsenkeller Dresden. From top to bottom: 61\,Bq calibration sample in Felsenkeller, 17\,mBq activated sample \#31 at the surface of the Earth, the same sample in Felsenkeller.}
 \label{fig:Offline_spectra}
\end{figure}

Using these calibration samples in exactly the same geometry as the activated sample, the $^{44}$Ti/$^{44}$Sc activity of the sample under study was simply determined from a comparison of the counting rates in the 78\,keV $^{44}$Ti and 1157\,keV $^{44}$Sc lines, with the true coincidence summing corrections canceling out (Fig.~\ref{fig:Offline_spectra}). Due to its short half life of 4\,hours, $^{44}$Sc is in secular equilibrium with its $^{44}$Ti mother both in the calibration and in the activation samples. The 68\,keV $^{44}$Ti peak was excluded from the comparison due to a low-energy shoulder by a lead fluorescence line that did not scale exactly with the $^{44}$Ti activity but was also affected by parasitic activities in the activated sample.

In addition to the expected $^{44}$Ti, the activation samples showed parasitic $^{22}$Na that had been produced by the $^{19}$F($\alpha$,n)$^{22}$Na reaction during the irradiation. It is not clear whether the fluorine was contained in the target material or in the backing. If one pessimistically assumes all the fluorine to be contained in the calcium hydroxyde and uses the known $^{19}$F($\alpha$,n)$^{22}$Na cross section \cite{Norman84-PRC}, based on the measured $^{22}$Na activity one arrives at a fluorine/calcium ratio inside the target of 0.005, changing the effective stopping power only negligibly. The Compton background from the $^{22}$Na lines did not limit the statistical precision of the activity determination. 

For comparison, one of the activated samples was also counted in a low-background HPGe setup at the Earth's surface, with a graded shield including a 15\,cm thick layer of lead. The $^{44}$Ti/$^{44}$Sc lines were entirely covered by cosmic-ray induced background in the surface-based spectrum (Fig.~\ref{fig:Offline_spectra}). The $^{44}$Ti/$^{44}$Sc activities measured for samples \#31 and \#32 lie in the mBq range (Table~\ref{table:Targets}).

\begin{table}
  \caption{Targets used for the $^4$He$^{++}$ irradiations. The $^{44}$Ti activities given refer to the end of the irradiation.}
  \begin{tabular}{lrr}
    \hline
    \hline
						& \multicolumn{1}{c}{\#31}	& \multicolumn{1}{c}{\#32} \\
    \hline
    Implanted $^4$He$^{++}$ dose [mC]		& 584				& 402 \\
    1083\,keV yield degradation			& ($8.2\pm1.6$)\%		& none \\
    Measured $^{44}$Ti activity [mBq]		& $17.0\pm0.6$			& $12.9\pm0.6$ \\
    \hline
    \hline
  \end{tabular}
\label{table:Targets}
\end{table}

\section{Results}

\begin{figure*}[]
 \includegraphics[width=1.0\textwidth]{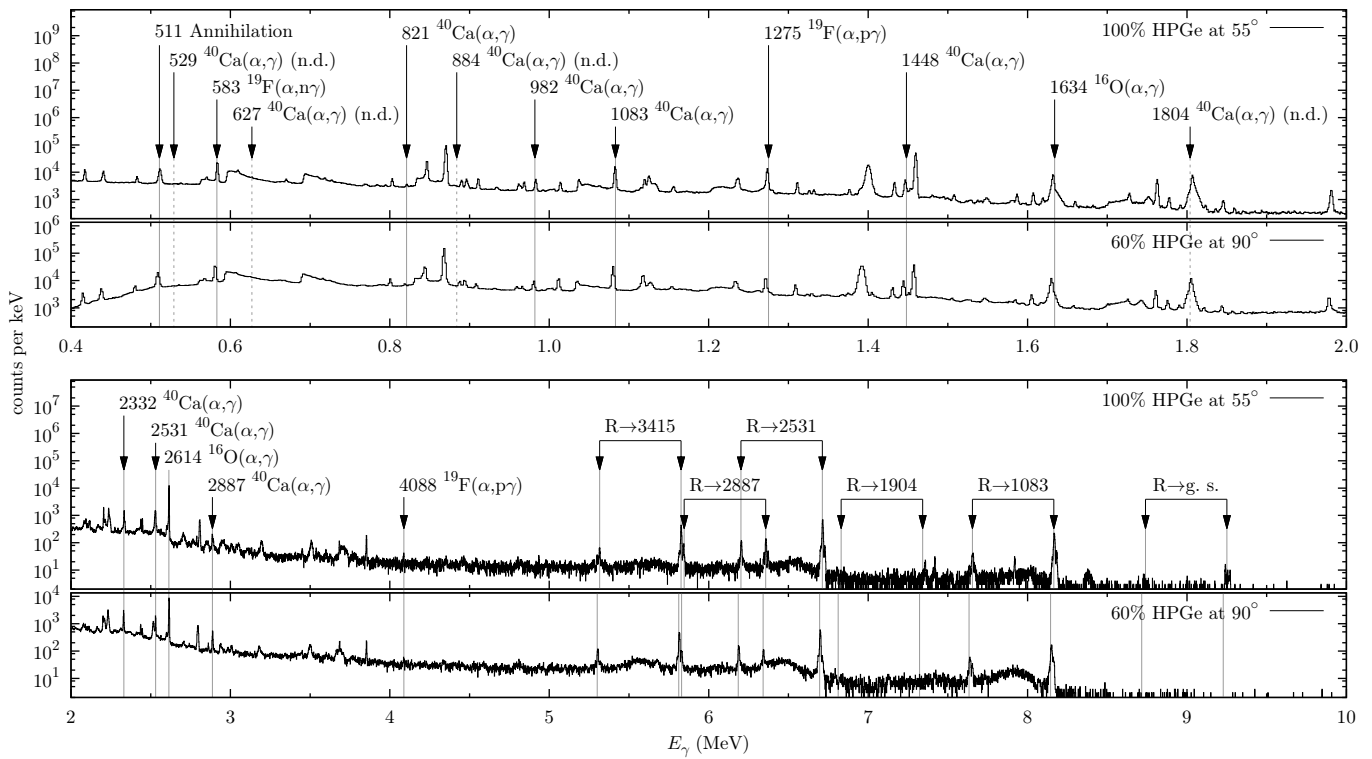}
 \caption{In-beam $\gamma$-ray spectrum for $E_\alpha = 4.5$\,MeV, target \#31. Above 5\,MeV, the primary $\gamma$ rays from the resonance triplet are seen, together with their single-escape lines. At lower energy, the secondary $\gamma$ rays from the reaction under study and also some contaminant $\gamma$ rays can be seen. Undetected $^{40}$Ca($\alpha$,$\gamma$)$^{44}$Ti secondary $\gamma$ rays have been marked with dashed lines and labeled as ''n.d.'' (not detected) in the plot.}
 \label{fig:Overall_gamma_spectrum}
\end{figure*}

The $\gamma$-ray spectra observed with the in-beam detectors are dominated by the $\gamma$ rays from the $^{40}$Ca($\alpha$,$\gamma$)$^{44}$Ti reaction under study here (Fig.~\ref{fig:Overall_gamma_spectrum}). The main background in the in-beam spectra is from $\alpha$-induced reactions on $^{19}$F that is contained either in the tantalum backing or in the Ca(OH)$_2$ target. In addition, some secondary $\gamma$ rays from the $^{16}$O($\alpha$,$\gamma$)$^{20}$Ne reaction have been identified.

\subsection{Decay branching ratios of the 4.5\,MeV triplet}

\begin{figure*}
 \includegraphics[width=1.0\textwidth]{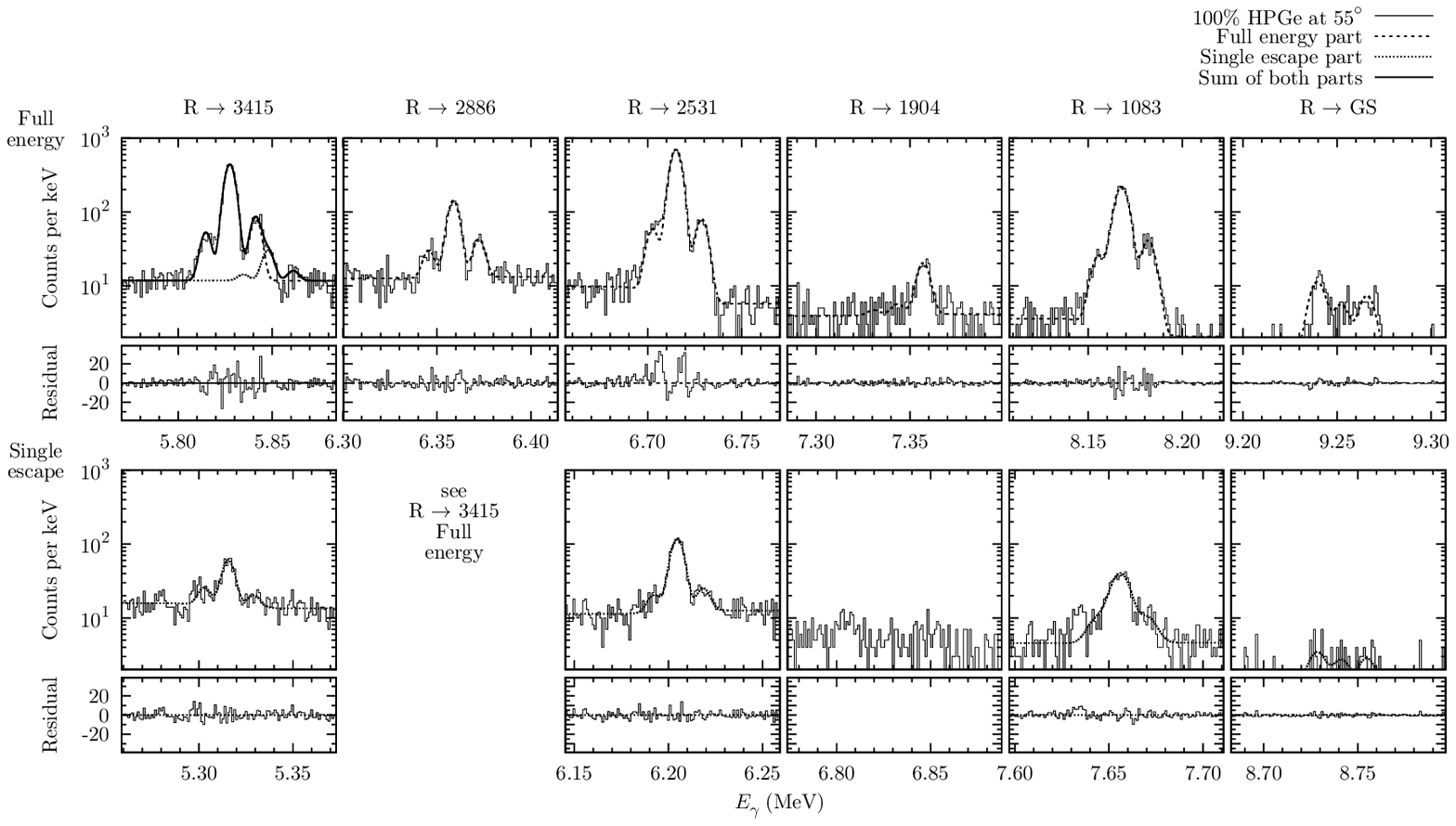}
 \caption{In-beam $\gamma$ rays of the resonance triplet at 4.5\,MeV. Top row: Primary $\gamma$ rays from the decay of the resonance triplet to several excited states in $^{44}$Ti. The triplet structure is apparent, but with different relative strengths due to the different branching ratios. Bottom row: Single escape peaks for the primary $\gamma$ rays from the top row. The single escape peak of the R$\rightarrow$2886 $\gamma$ ray is shown as the orange structure to the right of the R$\rightarrow$3415 full energy peak. Below each panel, the difference between observed spectrum and fit is shown. Level energies are given in keV.}
 \label{fig:triplet_100_31}
\end{figure*}

The decay branching ratios of the resonance triplet at 4.5\,MeV were determined from the in-beam $\gamma$-ray spectrum observed at 55$^\circ$. To this end, both the full energy and the single escape $\gamma$ rays were fitted (Fig.~\ref{fig:triplet_100_31}). 

The resulting branching ratios (Table~\ref{table:branching}) are mostly in good agreement with the literature \cite{Dixon80-CJP}. It can be seen that the branching ratio for the decay to the 1083\,keV first excited state in $^{44}$Ti is different in the present work: higher for the lowest-energy resonance and lower for the other two resonances. This level is fed by higher-lying states, so it is possible that the previous data were affected by uncertainties in the feeding subtraction. The present branching data depend only on the primary $\gamma$ rays, which are free from this effect. 

For the two higher-energy resonances, the new branching ratios are overall more precise than the previous ones. One problem in the present data is the decay to the 1904\,keV second excited state in $^{44}$Ti. The primary $\gamma$ ray for this decay is affected by inevitable Compton background from higher-energy peaks. An analysis based on the secondary $\gamma$ rays from the decay of the 1904\,keV level is in principle possible, but the significant feeding and branching corrections prevent a better precision than that of the literature value. 

Also from the fit, the distances between the three resonances have been determined, in the center-of-mass system: $12.4\pm0.2$\,keV between the lower-energy and the central resonance, and $13.7\pm0.2$\,keV between the central and the higher-energy resonance.

\begin{table*}[b!!]
  \caption{$\gamma$-ray branching ratios of the 4.5\,MeV resonance triplet, observed at 55$^{\circ}$ with respect to the beam axis. The present data are compared with those of Ref.~\cite{Dixon80-CJP}. For the 9227$\rightarrow$1904 and 9215$\rightarrow$1904 decays, the present data show only upper limits (see Fig.~\ref{fig:triplet_100_31}), so the previous results \cite{Dixon80-CJP} are adopted instead. Level energies are given in keV.}
  \begin{tabular}{D{.}{.}{0}D{.}{.}{1}@{$\,\pm\,$}D{.}{.}{1}@{\quad}D{.}{.}{2}@{$\,\pm\,$}D{.}{.}{2}@{\quad}D{.}{.}{1}@{$\,\pm\,$}D{.}{.}{1}
		  @{\quad}c@{\quad}
		  D{.}{.}{0}@{$\,\pm\,$}D{.}{.}{0}@{\quad}D{.}{.}{1}@{$\,\pm\,$}D{.}{.}{1}@{\quad}D{.}{.}{1}@{$\,\pm\,$}D{.}{.}{1}}
  \hline
  \hline
	& \multicolumn{6}{c}{Present work}		& \qquad\qquad	& \multicolumn{6}{c}{Dixon {\it et al.}~\cite{Dixon80-CJP}} \\
	& \multicolumn{6}{c}{Initial level (keV)}	&		& \multicolumn{6}{c}{Initial level (keV)} \\
	\cline{2-7}							\cline{9-14}
	& \multicolumn{2}{c}{9215}
			& \multicolumn{2}{c}{9227}
					& \multicolumn{2}{c}{9239}
							&		& \multicolumn{2}{c}{9215}
											& \multicolumn{2}{c}{9227}
													& \multicolumn{2}{c}{9239} \\
  \multicolumn{1}{c}{Final} \\
  \multicolumn{1}{c}{level} \\
  \multicolumn{1}{c}{(keV)}
	& \multicolumn{2}{c}{(\%)}
			& \multicolumn{2}{c}{(\%)}
					& \multicolumn{2}{c}{(\%)} \\
  \hline
     0	& 12.9	& 1.3	&  0.70	& 0.08	&  6.1	& 0.5	&		& 10	& 1	&  0.6	& 0.2	&  4.0	& 0.5 \\
  1083	& 20.1	& 2.3	& 21.4	& 0.5	& 23.6	& 1.4	&		& 16	& 2	& 23	& 2	& 27	& 2 \\
  1904	& \multicolumn{2}{c}{}
			& \multicolumn{2}{c}{}
					&  7.7	& 1.0	&		&  2	& 1	& \multicolumn{2}{c}{$<1$}
													&  5	& 1 \\
  2531	& 34	& 4	& 45.8	& 0.7	& 28.3	& 1.6	&		& 41	& 2	& 45	& 2	& 28	& 2 \\
  2886	&  9.4	& 1.7	&  7.7	& 0.3	& 11.0	& 0.8	&		& 11	& 2	&  8	& 2	& 11	& 2 \\
  3415	& 21.2	& 1.9	& 23.4	& 0.6	& 23.3	& 1.1	&		& 20	& 2	& 23	& 2	& 25	& 2 \\
  \hline
  \hline
  \end{tabular}
\label{table:branching}
\end{table*}
\subsection{Total resonance strength, activation method}
\label{subsec:TotalStrengthActivation}

For an infinitely thick target, the experimental yield $Y_{\infty}$ as a function of the resonance strength $\omega\gamma$ is given by the following relation~\cite{Iliadis07-Book}:
\begin{equation}\label{eq:Yield_omega_gamma}
Y_{\infty} = \frac{\lambda_{\rm res}^2}{2} \frac{\omega\gamma}{\varepsilon_{\rm eff}^{\rm He}(E_{\rm res})}
\end{equation}
where $\lambda_{\rm res}$ is the de Broglie wavelength at the resonance energy $E_{\rm res}$ and $\varepsilon_{\rm eff}^{\rm He}$ is the effective stopping power given by Eq.~(\ref{eq:EffectiveStoppingPower_He}). 

During the irradiation, the three resonances were activated together. As the square of the de Broglie wavelength varies only by 0.3\% between neighboring resonances of the triplet, much less than the experimental uncertainties, a total resonance strength formed by the sum of the three resonances can be computed (Table~\ref{table:TotalResonanceStrength}). 


Due to the finite energetic thickness of the target, about 42\,keV for 4.5\,MeV $\alpha$ particles, a finite target thickness correction has to be considered when using the thick target yield formula Eq.\,\ref{eq:Yield_omega_gamma}. In order to calculate this correction, an upper limit for the previously unknown $\alpha$ width of the central resonance is determined here. Using a CaO target called \#12 with a nominal thickness of 12\,$\mu$g/cm$^2$ for a scan of the 4.5\,MeV resonance triplet, for the $\alpha$ width a 1$\sigma$ (2$\sigma$) upper limit of $\Gamma_\alpha$ $<$ 0.6\,keV (0.9\,keV) has been determined for the central resonance. The 1$\sigma$ upper limit was then used for the finite target thickness correction for all three resonances. 
The significant energy straggling of the $\alpha$ particles inside the target, 10\,keV for the full thickness of targets \#31 and \#32, also has to be taken into account.

Taking into account the irradiation energy, the true form of the yield curve \cite{Iliadis07-Book}, and the energy straggling \cite{Ziegler10-NIMB}, a correction factor of $0.994\pm0.001$, $0.982\pm0.004$, and $0.63\pm0.07$, respectively, for the upper, middle, and lower resonance has been calculated. Note the correction becomes truly significant only for the weakest resonance that is activated in the deeper layers of the target and contributes less than 10\% to the total strength. The finite target thickness correction affects the summed resonance strength by ($5.5\pm1.1$)\%, again assuming a conservative relative uncertainty of 20\%. 

The major contribution to the error budget in the resonance strength (Table~\ref{table:Errors}) is the target stoichiometry. Using the adopted stoichiometry of Ca(OH)$_{1.88\pm0.21}$, the error on the oxygen and hydrogen content propagates to 6.0\% uncertainty in the effective stopping power in Eq.~(\ref{eq:EffectiveStoppingPower_He}). The normalization of the stopping power \cite{Ziegler10-NIMB} contributes another 1.9\% uncertainty. The activity of the $^{44}$Ti/$^{44}$Sc calibration samples is known to 1.5\%~\cite{Schmidt11-Diplom}, and the counting statistics in Felsenkeller contributes 3.1\% (4.2\%) uncertainty for target \#31 (\#32), respectively; hence 2.5\% for both targets combined.

\begin{table}
  \caption{Error budget for the total resonance strength, for the activation and for the in-beam $\gamma$-spectroscopy methods.}
  \begin{tabular}{lrrr}
    \hline
    \hline
								& Activ.	& Common	& In-beam \\ \hline
    Target stoichiometry					&		& 6.0\%	& \\
    Stopping power \cite{Ziegler10-NIMB}			&		& 1.9\%		& \\
    Beam current						&		& 1.0\%		& \\
    Target degradation (target \#31)				& 		& 1.6\%		& \\
    Finite target thickness 					&  		& 1.1\%		&  \\
    $\gamma$-ray detection efficiency \cite{Schmidt11-Diplom}	& 1.5\%		&		& 3\% \\
    $^{44}$Ti half-life \cite{Ahmad06-PRC}			& 0.5\%		&		& \\
    $\gamma$-ray angular distribution	 \cite{Dixon80-CJP}	& 		&		& 10\% \\
    \hline
    Total systematic error					& 1.6\%		& 6.7\%		& 10\% \\
    Statistical error						& 2.5\%		&		& 1.0\% \\
    \hline
    \hline
  \end{tabular}
  \label{table:Errors}
\end{table}

\subsection{Total resonance strength, in-beam $\gamma$ spectroscopy method}
\label{subsec:TotalStrengthInBeam}

In spite of the complicated decay scheme, a determination of the resonance strengths from the in-beam $\gamma$-ray spectra is attempted here. 

The angular distributions for the primary and secondary $\gamma$ rays from the present resonance triplet have been studied previously, using Ge(Li) detectors~\cite{Dixon80-CJP}. The distribution was found to be anisotropic in all cases except for the isotropic 1904$\rightarrow$1083 (0$^+$$\rightarrow$$2^+$) line at 821\,keV. However, this line is not very strong due to its weak branching and is located in a region of high $\gamma$ continuum, so it is not attempted to use it for a resonance strength measurement. 

Instead, the high-energy primary $\gamma$ rays are used. Because of the strongly anisotropic nature of several transitions, the analysis is again limited to the 55$^\circ$ detector. For this detector, the corrections for the angular distribution~\cite{Dixon80-CJP} amount to less than 11\% in all cases, except for the weak transition to the ground state, where the observed yield had to be corrected down by 40\% in order to take into account the strong anisotropy. For the weak transition to the 1904\,keV excited state, no angular distribution data exist, and isotropy was assumed here. 

\begin{table}
  \caption{Total resonance strength of the 4.5\,MeV resonance triplet. The first uncertainty given is the statistical uncertainty. For the average values for one given method, in addition also the part of the systematic uncertainty pertinent to this method (columns 2 and 4 in Table~\ref{table:Errors}) is listed. The systematic uncertainty common to both methods (column 3 in Table~\ref{table:Errors}) was taken into account for the global average of both methods.}
  \begin{tabular}{llr@{$\,\pm\,$}l}
    \hline
    \hline
    Target	& Method			& \multicolumn{2}{l}{$\omega\gamma$ [eV]} \\
    \hline
    \#31	& Activation			& 8.6	& 0.3$_{\rm stat}$	 \\	
    \#32	& Activation			& 8.3	& 0.3$_{\rm stat}$	 \\
    \multicolumn{2}{l}{Average, activation}	& 8.5	& 0.2$_{\rm stat}\pm0.1_{\rm syst}^{\rm act}$ \\
    \hline
    \#31	& In-beam			& 7.6 	& 0.2$_{\rm stat}$ \\
    \#32	& In-beam			& 7.3	& 0.1$_{\rm stat}$ \\
    \multicolumn{2}{l}{Average, in-beam}	& 7.4	& 0.1$_{\rm stat}\pm0.8_{\rm syst}^{\rm in\text{-}beam}$ \\
    \hline
    \multicolumn{2}{l}{Global average}		& 8.4	& 0.6 \\
    \hline
    \hline
  \end{tabular}
  \label{table:TotalResonanceStrength}
\end{table}

The total resonance strength data obtained by the activation and by the in-beam $\gamma$-spectroscopy method are compared in Table~\ref{table:TotalResonanceStrength}. The in-beam data are on average 13\% lower than the activation data. This difference is probably due to the strongly anisotropic angular distribution, which has been studied experimentally by just one group~\cite{Dixon80-CJP}. No details are given in that work on experimental aspects of the angular distribution measurements, such as the number of detectors used simultaneously and the geometry of the target chamber. An earlier $^{40}$Ca($\alpha$,$\gamma$)$^{44}$Ti angular distribution measurement by the same group included one Ge(Li) detector as a yield monitor and a second Ge(Li) at a target distance of just 4\,cm for the measurement~\cite{Simpson71-PRC}. It seems prudent to assume a 10\% uncertainty for the angular corrections derived on the basis of these previous works~\cite{Simpson71-PRC,Dixon80-CJP}.

Taking into account which uncertainties are common and which are individual to the data point and method used, a global average for the total strength of the triplet of 8.4$\pm$0.6\,eV is found (tables~\ref{table:TotalResonanceStrength} and~\ref{table:LiteratureReview}).
 
\subsection{Strengths of individual resonances of the triplet}

From an experimental point of view, the triplet can be treated as one composite resonance due to the negligible difference in the squared de Broglie wavelength. For an astrophysical scenario, however, the situation may be different. The astrophysical reaction rate for a narrow resonance is given by~\cite{Iliadis07-Book}
\begin{equation}\label{eq:TNRR}
N_{\rm A} {\langle}\sigma v{\rangle} = N_{\rm A} \left( \frac{2\pi}{\mu k_{\rm B}T} \right)^{\frac{3}{2}} \hbar^2 \exp\left(-\frac{E_{\rm r}}{k_{\rm B}T}\right) \omega \gamma
\end{equation}
with $\mu$ the reduced mass of the reaction partners, $k_{\rm B}$ the Boltzmann constant, and $T$ the temperature of the astrophysical plasma. $N_{\rm A} {\langle}\sigma v{\rangle}$ depends exponentially on $E_{\rm r}/(k_{\rm B}T)$. For $T = 4$\,GK, where the present resonance triplet starts dominating the astrophysical reaction rate, the 24\,keV difference in resonance energies between the lower and the higher resonances of the triplet leads to a 7\% change in $\exp(-E_{\rm r}/k_{\rm B}T)$, comparable with the present experimental uncertainties.  

Therefore it is useful to understand how much strength each resonance of the triplet contributes individually. Based on the in-beam data and on the total resonance strength (Table~\ref{table:TotalResonanceStrength}), the individual strengths for the three resonances included in the triplet have been computed (Table~\ref{table:triplet_resonance_strenghts}). The strength for the lowest resonance in the triplet is significantly larger than the previous value \cite{Dixon80-CJP}, but the ordering of the resonance strengths, with the lower resonance the weakest and the central resonance the strongest, remains intact. 

\begin{table}
  \caption{Individual strengths of the three resonances of the 4.5\,MeV triplet in eV, from the present work and from Ref.~\cite{Dixon80-CJP}.}
  \begin{tabular}{lD{.}{.}{2}@{$\,\pm\,$}D{.}{.}{2}D{.}{.}{2}@{$\,\pm\,$}D{.}{.}{2}D{.}{.}{2}@{$\,\pm\,$}D{.}{.}{2}}
    \hline
    \hline
    $E_\alpha$ (keV)		& \multicolumn{2}{c}{4497}	& \multicolumn{2}{c}{4510}	& \multicolumn{2}{c}{4523} \\
    $E_{\rm x}$ in $^{44}$Ti (keV) & \multicolumn{2}{c}{9215}	& \multicolumn{2}{c}{9227}	& \multicolumn{2}{c}{9239} \\
    \hline
    Present work		& 0.92	& 0.20			& 6.2	& 0.5			& 1.32	& 0.24 \\
    Dixon {\it et al.} \cite{Dixon80-CJP}	& 0.5	& 0.1			& 5.8	& 1.2			& 2.0	& 0.4 \\
    \hline
    \hline
  \end{tabular}
  \label{table:triplet_resonance_strenghts}
\end{table}

\subsection{Yield of the 1083\,keV line relative to $^{44}$Ti production}
\label{subsec:Yield1083}

In order to assist future experiments, the number of 1083\,keV $\gamma$ rays emitted at 55$^\circ$ angle has been determined from the present in-beam $\gamma$ ray spectra and related to the number of $^{44}$Ti nuclei produced, as measured by the activation method. 
For target \#31 (\#32), a ratio of $N_{1083}^{55^\circ}$/$N_{^{44}{\rm Ti}}$ = 0.633$\pm$0.024 (0.638$\pm$0.031) is found, respectively. Averaging these two results, the following purely experimental value is found:

\begin{equation}\label{eq:1083ratio}
\frac{N_{1083}^{55^\circ}}{N_{^{44}{\rm Ti}}} = 0.635	\pm 0.021
\end{equation}

The error bar for this ratio includes also the uncertainty introduced by the finite target thickness corrections. The 55$^\circ$ germanium detector is at sufficiently far geometry that summing out corrections are below 1\% and can be neglected. 

It should be stressed that the ratio (\ref{eq:1083ratio}) is only valid at 55$^\circ$ angle; due to the significant angular distribution of this line \cite{Simpson71-PRC} it should not be used at other angles. Also, in order for the ratio to be applicable it is necessary that all three resonances are activated together.

\section{Discussion and outlook}

The $E_\alpha = 4.5$\,MeV resonance triplet in the $^{40}$Ca($\alpha$,$\gamma$)$^{44}$Ti reaction has been studied in a precision experiment, using both the activation and the in-beam $\gamma$ spectroscopy methods. The target composition has been studied both by the ERD method and by nuclear reactions. For each resonance of the triplet, the individual strength and the decay branching ratios have been re-determined. The $\alpha$ width of the central resonance has been determined.

\begin{figure}[t!!]
 \includegraphics[width=1.0\columnwidth]{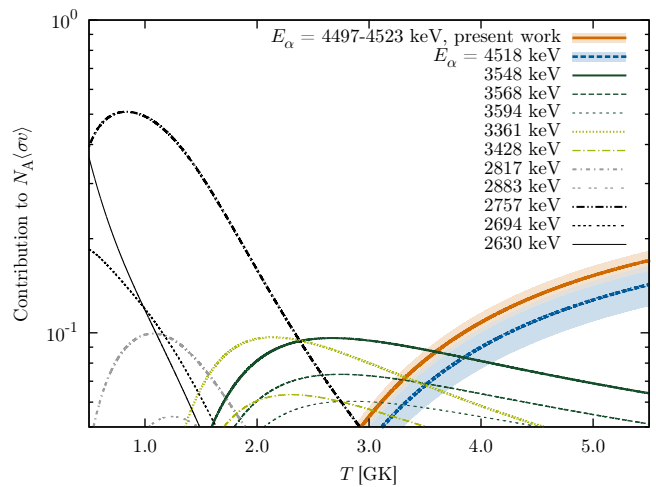}
 \caption{(Color online) Relative contributions by the individual resonances to the total astrophysical reaction rate. The present new strengths for the three 4.5\,MeV resonances replace the strength reported by Ref.~\cite{Vockenhuber07-PRC} at $E_\alpha=(4518\pm22)$\,keV. For the other resonances, the previous energies and strengths~\cite{Vockenhuber07-PRC} are used.}
 \label{fig:TNRR_Vock}
\end{figure}

The present results for the triplet resonance strength are more precise than, but within uncertainties consistent with previous work using in-beam $\gamma$-spectroscopy~\cite{Dixon80-CJP,Robertson12-PRC}, accelerator mass spectrometry~\cite{Nassar06-PRL}, and recoil separator~\cite{Vockenhuber07-PRC} methods. The new, precise strength value and branching ratios may be used as reference values in future work investigating lower-energy resonances. 

The present new data are limited to the $E_\alpha = 4.5$\,MeV resonance triplet; therefore it is not attempted here to re-determine the $^{40}$Ca($\alpha$,$\gamma$)$^{44}$Ti astrophysical reaction rate. Such a re-determination would require similarly precise information also on the lower-energy resonances. 

However, in order to illustrate the impact of the present work alone, the relative contributions of the various $^{40}$Ca($\alpha$,$\gamma$)$^{44}$Ti resonances to the astrophysical reaction rate in Eq.~(\ref{eq:TNRR}) have been plotted (Fig.~\ref{fig:TNRR_Vock}). It is clear that the 4.5\,MeV triplet, with its present slightly enhanced strength, plays a strong role for temperatures $T$ $>$ 3.5\,GK, towards the higher end of the relevant temperature range~\cite{Hoffman10-ApJ} for the $\alpha$-rich freezeout. 

The astrophysically interesting question at the outset of the present work is whether the assumed reaction rate for $^{44}$Ti nucleosynthesis is correct or not. For the higher temperature range of the $\alpha$-rich freezeout it can be stated that there are no major surprises. A thermonuclear reaction rate taking the present new data into account would be slightly higher than Ref.~\cite{Vockenhuber07-PRC} and slightly lower than Ref.~\cite{Robertson12-PRC}, the two most recent works on this topic. 

Some surprises may lie, however, at lower energies $E_\alpha<3$\,MeV, where there are only upper limits~\cite{Nassar06-PRL,Robertson12-PRC} or data with large error bars~\cite{Dixon80-CJP,Vockenhuber07-PRC} for the resonance strengths. It is planned to extend the present experiment to lower energies in the future, using the 4.5\,MeV resonance triplet as point of reference.

\begin{acknowledgments}
Technical support by the staff and operators of the HZDR ion beam center, by Andreas Hartmann (HZDR), by Bettina Lommel and the GSI target laboratory is gratefully acknowledged. --- This work was supported in part by DFG (BE4100/2-1), by the Helmholtz Association Nuclear Astrophysics Virtual Institute (NAVI, VH-VI-417), by EuroGENESIS, by the European Union (FP7-SPIRIT, contract no. 227012, and ERC StG 203175), and by OTKA (K101328).
\end{acknowledgments}

\end{document}